\documentclass[fleqn,twoside]{article}
\usepackage{espcrc2}

\usepackage{graphicx}
\usepackage[figuresright]{rotating}

\newcommand{\AmS}{{\protect\the\textfont2
  A\kern-.1667em\lower.5ex\hbox{M}\kern-.125emS}}

\title{\vspace*{-5cm} \hbox to \hsize{\hfill \vbox{\hsize=6cm \noindent \small \it 
The Restless High-Energy Universe\\
Proceedings of the symposium dedicated\\
to six years of successful\\
BeppoSAX operations\\
Amsterdam, May 5-8, 2003}}\vspace*{3cm}
The Puzzles of RX J1856.5-3754: Neutron Star or Quark Star?}

\author{J. E. Tr\"{u}mper\address[MPE]{Max-Planck-Institut 
                                       f\"{u}r extraterrestrische Physik, 
                                     D-85741 Garching, Germany}, 
        V. Burwitz\addressmark[MPE],  
        F. Haberl\addressmark[MPE], and  
        V.E. Zavlin\addressmark[MPE]}

\begin{document}

\begin{abstract}

We discuss recent Chandra and XMM-Newton observations of the 
bright isolated neutron star RX\,J1856.5$-$3754 and suggest that 
the absence of any line features is due to effects of a high 
magnetic field strength ($\sim\,10^{13}$\,G). Using different models for 
the temperature distribution across the neutron star surface 
assuming blackbody emission to fit the optical and X-ray 
spectrum and we derive a conservative lower limit of the 
''apparent`` neutron star radius of 16.5\,km\,$\times$\,(d/117\,pc). 
This corresponds to the radius for the ''true`` (de-redshifted) 
radius of 14\,km for a 1.4\,M$_{\odot}$ neutron star, indicating a 
stiff equation of state at high densities. A comparison of the 
result with mass-radius diagrams shows that quark stars and 
neutron stars with quark matter cores can be ruled out with 
high confidence.
\vspace{1pc}
\end{abstract}

\maketitle

\section{INTRODUCTION}
The excellent soft X-ray response of ROSAT pointed and all-sky 
observations \cite{Tru83} have led to the discovery of eight thermally 
emitting objects (listed in Table~\ref{table:1}) having roughly the size 
of a neutron star but showing not any radio emission\,\cite{Hab03}.

Among these objects RX\,J1856.5-3754 is the brightest and thus 
best qualified for detailed studies aiming at a determination 
of its radius, surface gravity and gravitational redshift by 
means of X-ray spectroscopy, allowing to determine the equation 
of state of matter at supranuclear densities. 
In this paper, we concentrate primarily on this object and first
summarize the main observational facts. 

RX\,J1856.5$-$3754 (henceforth RXJ1856) was discovered serendipitously
in a ROSAT field by Walter et al.\,\cite{Wal96}. Using the HST 
Walter \& Matthews\,\cite{Wal97} identified the X-ray source with 
a faint blue star (V\,$\sim$\,26\,mag). Its distance and proper 
motion were measured with HST as well, yielding (117\,$\pm$\,12)\,pc 
and 0.33\,arcsec/year, respectively \cite{Wal02,Neu01}. 
With the VLT van Kerkwijk \& Kulkarni\,\cite{Ker01} found a faint nebula 
surrounding the point source which has a cometary-like geometry 
with a 25" tail extending along the direction of motion. 
None of the X-ray observations revealed any variability 
on time scales up to ten years. 
The so far best upper limit of 1.3\%\,(2$\sigma$) on periodic 
variations in the range $10^{-3}$\,$-$\,50\,Hz has been established 
by Burwitz et al.\,\cite{Bur03} using a XMM-Newton EPIC-pn observation. 
Chandra LETG observations with high spectral resolution show 
a featureless spectrum that can be fit by a Planckian  spectrum 
with a temperature of 63\,$\pm$\,3\,eV \cite{Bur03}. Compared with the optical 
spectrum which shows a Rayleigh-Jeans slope, the X-ray spectrum 
is reduced by a factor of $\sim$\,6. Therefore the overall spectrum of 
the source has often been described by a two-temperature 
blackbody model e.g. \cite{Bur03,Pon02,Pav03,Pav02}.

A large number of papers have been dealing with the nature 
of this compact object and the proposed interpretations 
include almost everything which has ever been discussed in 
this context, ranging from ''normal`` neutron stars with a stiff 
or soft equation of state over neutron stars having a quarks core 
to bare strange quark stars. 
In this paper, we show that the observational data require a 
normal neutron star with a rather stiff equation of state.

\section{THE MAGNETIC FIELD STRENGTH OF RX\,J1856.5$-$3754}

The lack of any significant spectral features in the LETG 
spectrum excludes magnetic fields of (1.3\,$-$\,7)\,$\times$\,10$^{11}$\,G 
(for electrons) and (0.2\,$-$\,1.3)\,$\times$\,10$^{14}$\,G (for protons), 
see Burwitz et al. \cite{Bur01}. 
This leaves the possibility of a low magnetic field 
characteristic for millisecond pulsars or a high magnetic 
field typical for normal pulsars open. 
Unfortunately, due to the absence of a periodicity the 
usual estimate of the magnetic field of RXJ1856 based on 
the rotating dipole model is not possible. 
Using phenomenological arguments based on the very small 
pulsed fraction in X-rays and on a comparison with other 
objects van Kerkwijk \& Kulkarni \cite{Ker01} have argued that the 
star has a relatively low magnetic field of a few 10$^{11}$\,G 
which may be marginally consistent with the absence of 
proton cyclotron lines. But this is not the only possibility. 
We estimate the magnetic field making use of the spin-down 
luminosity 
dE/dt\,$\sim$\,4\,$\times$\,10$^{32}$\,erg/s 
required for powering the 
cometary-like emission nebula \cite{Ker01} and of the age of the 
star (t\,$\sim$\,5\,$\times$\,10$^{5}$ years) inferred from its proper motion 
and the distance to its likely birthplace in the Upper 
Sco OB association \cite{Wal02}: assuming as usual the validity 
of magnetic dipole braking we find a period of a $\sim$1.8\,sec 
and a magnetic field strength of  $\sim$\,1.1\,$\times$\,10$^{13}$G. 
We emphasise that these figures are very similar to those 
of the second brightest object of this kind, the pulsating 
source RX\,J0720.4$-$3125 (henceforth RXJ0720) whose spectral 
characteristics are also very similar to those of RXJ1856 
(see below). 
While the estimate of dE/dt may be considered as rather 
reliable, the age derived from the birthplace argument 
is not so certain. However, an age of t\,$\sim$\,5\,$\times$\,10$^{5}$\,years 
(with an uncertainty of a factor of two) is fully consistent 
with what we know empirically about the cooling of neutron 
stars. 
We therefore conclude that the magnetic field of RXJ1856 
is probably large, i.e. of the order of $>\,10^{13}$ G. 
To support this, it will be important to exclude the 
alternative hypothesis of a millisecond pulsar \cite{Ker01,Pav03} 
by means of a high time resolution observation with 
XMM-Newton which has already been approved.
\begin{table*}[tbh]
\caption{Parameters of X-ray-dim isolated neutron stars.}
\label{table:1}
\renewcommand{\arraystretch}{1.1} 
\begin{tabular}{lcccccl}
\hline

&&&&&&\\[-4.5mm]
Source Name	        &  P  & \.{P}              & L$_{\mbox{x}}$               & ${\mbox{kT}}_{\mbox{BB}}$ &	 d  &	Opt.    \\
	                &  (s) & (s\,s$^{-1}$)             & (erg\,s$^{-1}$)        &      (eV) &	(pc)&	(mag)    \\
\hline
RX J0420.0$-$5022$^{a}$ &   3.45 & $-$			  & 2.7$\times 10^{30}$   &	44 &	100 &	B\,$>$\,25.5 \\
RX J0720.4$-$3125	&   8.39 & (3$-$6)$\times 10^{-14}$ & 2.6$\times 10^{31}$ &	85 &	100 &	B\,$=$\,26.6 \\
RX J0806.4$-$4123	&  11.37 & $-$  		  & 5.7$\times 10^{30}$   &	95 &	100 &	B\,$>$\,24   \\
1RXS J130848.6+212708	&  10.31 & $<$6$\times 10^{-12}$  & 5.1$\times 10^{30}$   &	90 &	100 &	${\mbox{m}}_{\mbox{50CCD}}$\,=\,28.6 \\
RX J1605.3+3249		&  $-$	 & $-$			  & 1.1$\times 10^{31}$   &	92 &	100 &	B\,$>$\,27   \\
RX J1836.2+5925		&  $-$	 & $-$			  & 5.4$\times 10^{30}$   &	43 &	400 &	V\,$>$\,25.2 \\
RX J1856.5$-$3754	&  $-$	 & $-$			  & 1.5$\times 10^{31}$   &	63 &	117 &	V\,$=$\,25.7 \\
1RXS J214303.7+065419	&  $-$	 & $-$			  & 1.1$\times 10^{31}$   &	90 &	100 &	R\,$>$\,23   \\
\hline
\end{tabular}
$^{a}$\,Period and temperature from XMM-Newton data (Haberl et al. {\it in preparation}).
\end{table*}

\section{THE FEATURELESS X-RAY\\ SPECTRUM OF RX\,J1856.5$-$3754}

The main puzzle of RXJ1856 is the obser-vational fact that its 
X-ray spectrum is completely featureless. 
It has been pointed out by Burwitz et al.\,\cite{Bur03,Bur01} that 
nonmagnetic photospheric spectra assuming a pure iron 
composition are incompatible with the measured spectrum 
because the predicted Fe-L features are not detected 
with high significance. 
Even a solar composition model with its small abundance 
of metals leads to unacceptable spectral fits. 
Doppler smearing of the spectral lines due to fast rotation 
does not wash away completely the strongest spectral 
features \cite{Bra02,Pav02}. 
On the other hand hydrogen or helium photospheres can 
be excluded, because they over-predict the optical flux 
by a large factor \cite{Pav96}. Therefore any nonmagnetic 
photosphere can be firmly excluded. 
	
As a remedy it has been proposed that the star has no 
atmosphere but a condensed matter surface \cite{Bur01,Tur03}. 
Such a surface is expected to be reflective in the 
X-ray domain \cite{Tru78,Bri80} which could help to explain the 
low X-ray/optical flux ratio. 
\begin{figure*}[tbh]
\centerline{\includegraphics[width=11.8cm]{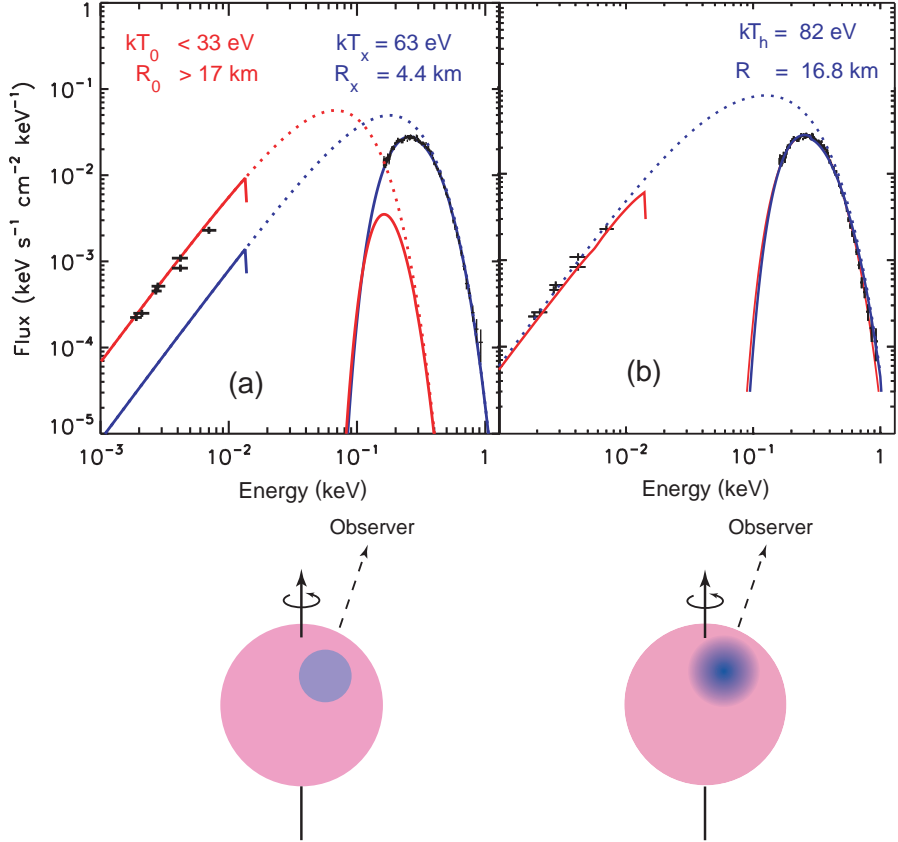}}
\vspace*{-0.7cm}
\caption{Blackbody fits to the optical and X-ray spectra of RX J1856.5-3754 
for a two-component model 
(a) and a model with a continuous temperature distribution (b).}
\label{fig:1}
\end{figure*}
Condensation of surface matter requires low tem-peratures 
and strong magnetic fields. To condense hydrogen at a 
temperature of kT\,=\,63\,eV a magnetic field of 5\,$\times\,10^{13}$\,G 
is required \cite{Lai01}. 
For iron it is not clear whether a condensate can exist at all. 
According to Lai\,\cite{Lai01} the cohesive energy of iron is uncertain, 
but condensation may occur at 3\,$\times\,10^{11}$\,G (at kT\,=\,63\,eV) while 
Neuhauser et al.\,\cite{Neu87} conclude that iron cannot condensate. 
Another problem is that the optical properties (the reflectivity 
depending on photon energy, polarisation and magnetic field angle) 
of a condensed matter surface have not been calculated taking 
into account the effects of atomic energy levels and of the 
solid state structure. 
Simplified models which calculate the reflectivities of a 
highly magnetized, high density electron plasma have been 
performed by Lenzen \& Tr\"{u}mper\,\cite{Tru78}, Brinkmann\,\cite{Bri80}, 
Zane et al.\,\cite{Zan03} and Turolla et al.\,\cite{Tur03}. 
They yield X-ray reflectivities of typically 10\,$-$\,30\,\% which 
smoothly depend on frequency. Close to the cyclotron frequency 
$\sim$\,50\,\% is reached. 
We note that this is not sufficient to fully explain the reduced 
X-ray flux which would require a reflectivity of $\sim$\,85\,\%.

Another possibility to suppress atomic line features 
is to employ super-strong magnetic fields. 
The energy levels for iron atoms in strong magnetic 
fields (10$^{13}$\,G) calculated by Neuhauser et al.\,\cite{Neu87} 
are pretty much smeared over the entire LETG energy 
range and have a spacing of 50\,$-$\,100\,eV. 
That could be easily resolved by the LETG (resolution $<$\,1\,eV). 
However, the individual levels show a B$^{0.4}$-dependence 
and therefore a smearing is expected to take place if 
the flux is integrated over the whole stellar surface. 
For a dipolar field with a factor of two variation of the 
magnetic field between pole and equator the lines would 
be broadened by 80\,$-$\,300\,eV . 
Thus it is plausible, that the combination of a dense 
level structure of the magnetic atoms with a dispersion 
of the magnetic field produces a spectrum which appears 
as a continuum seen with the LETG. 
We believe that this is the most promising model for 
explaining the featureless X-ray spectrum, but it has 
to be confirmed by detailed calculations.

\section{THE ABSENCE OF PERIODIC\\ VARIATIONS OF RX\,J1856.5$-$3754}

The upper limit of 1.3\,\% for the amplitude of periodic 
variations requires a pretty good alignment between 
the rotational axis and the line of sight if the 
overall spectrum is represented by a two-temperature 
blackbody model \cite{Bra02}. 
We note that this constraint can be somewhat (but not fully) 
relaxed if the X-ray flux were reduced due to reflection 
effects because the size of the X-ray emitting spot would 
be increased. 
Another possibility is that the rotational frequency 
of the neutron star is larger than the present limit 
of observations (50\,Hz). 
This will be checked by a forthcoming accepted XMM-Newton 
observation with the EPIC pn-CCD camera in using its high 
time resolution mode.

\section{COMPARISON WITH RELATED\\ OBJECTS}

(1) RX\,J0720.4-3125: The second brightest of the sources 
listed in Table 1 is RXJ0720 which looks like a twin of 
RXJ1856 in many respects: It was discovered in the ROSAT 
all sky survey and has a (blackbody) temperature of 
kT\,=\,79\,$\pm$\,4\,eV. 
The X-ray source was optically identified with a 
B\,=\,26.5\,mag object by Motch \& Haberl \cite{Mot98} and 
Kulkarni \& van Kerkwijk \cite{Kul98}. 
The optical to X-ray flux ratio is about 5. 
RXJ0720 shows X-ray pulsations with a period of 8.3914\,sec 
and a pulsed fraction of $\sim$\,10\,\% \cite{Hab97}. 
A timing analysis including all available archival and new ROSAT,
Chandra, and XMM-Newton data indicates that the source is 
spinning down with a rate of  dP/dt\,$\sim$\,10$^{-14}$\,s\,s$^{-1}$ \cite{Zan02}. 
When interpreted in terms of magnetic braking this results 
in a magnetic field of B\,$\sim$\,2\,$\times$\,10$^{13}$\,G.

A broad spectral feature at $\sim$\,270\,eV was found in the 
EPIC-pn-spectra and when interpreted as a proton or iron 
cyclotron resonance yields a magnetic field strength of 
$\sim$\,5\,$\times$\,10$^{13}$\,G and $<$\,2.5\,$\times$\,10$^{13}$\,G, 
respectively, consistent with the value derived from the spin down \cite{Hab03a}. 
	
(2) 1RXS\,J130848+2127 (=RBS1223): A strong broad band 
absorption feature was detected in the spectrum of this 
10.31\,s pulsar, with a similarly inferred magnetic field 
strength of (2\,$-$\,6)\,$\times$\,10$^{13}$\,G \cite{Hab03b}. 
This pulsar shows a double peaked pulse profile. 
	
The similarities of these three sources strengthen the 
case of a strong magnetic field of RXJ1856. 
Actually the main difference between them may be in the 
configuration of spin axis, magnetic axis and the line of sight.

\section{MODIFICATION OF THE\\ PLANCKIAN SPECTRUM}

Yet another similarity between these sources regards the 
shape of the X-ray continuum. 
Although in the cases of RXJ1856 and RXJ0720 a Planckian 
distribution yields a very good fit to the data a modified 
Planckian of the type E$^a$\,$\times$\,Planckian fits 
even somewhat better. 
This has been first discussed by Burwitz et al. \cite{Bur01} 
for RXJ1856 for which $a\,=\,1.28\,\pm\,0.30$, but 
also applies to RXJ0720 which has $a\,=\,0.98\,\pm\,0.30$. 
In compensation for the steeper continuum the interstellar 
column densities n$_{\mbox{H}}$  and temperatures of RXJ1856 are reduced 
by $\sim$\,50 and $\sim$\,10\,\%, respectively. 
We note that such a broad band modification does not affect 
the discussion of the absence of (narrow band) line features. 
But it may be connected with an energy dependent reflection 
coefficient of the emitting surface.

\section{A LOWER LIMIT TO THE RADIUS OF RX\,J1856.5$-$3754}  

\begin{figure*}[tbh]
\centerline{\includegraphics[width=11.5cm]{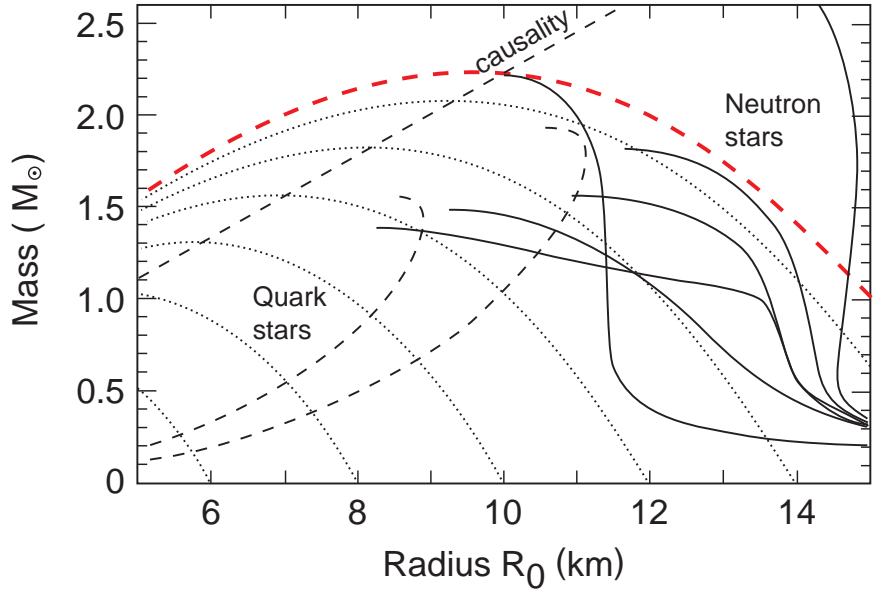}}
\vspace*{-0.7cm}
\caption{The mass-radius relations for various equations of state 
for the nuclear matter according to \cite{Pon02}. The thick dashed curve (red) 
represents the apparent neutron star radius derived from both the 
two-component and continuous temperature blackbody models.}
\label{fig:2}
\end{figure*}In the absence of any spectral features the determination 
of the neutron star radius hinges on the distance, for which 
we adopt d\,=\,117\,pc \cite{Wal02,Neu01}. 
In the following we first use the parameters of a two component 
blackbody model for the optical and X-ray spectrum of RXJ1856 
(Burwitz et al. \cite{Bur03}) which is shown in Fig. 1a. 
The blackbody radius and temperature of the X-ray emitting 
hot spot derived from the Chandra LETG spectrum are 
R$_{\mbox{x}}$\,=\,4.4\,km and kT$_{\mbox{x}}\,=\,63$\,eV, respectively. 
The optical spectrum is interpreted as the sum of the Rayleigh-Jeans 
spectra of both the hot and the cool component.  
This fixes 
(R${}_{\mbox{opt}}$)${}^2$\,$\times$\,T$_{\mbox{opt}}$\,+\,(R$_{\mbox{x}}$)${}^2$\,$\times$\,T${}_{\mbox{x}}$\,=\,33\,eV\,$\times$\,(17\,km)$^2$. 
The condition that the optical spectrum of the cool component 
does not show up as a deviation in the X-ray spectrum limits 
the corresponding temperature to kT$_{opt}$\,$<$\,33\,eV 
at the 3\,$\sigma$ level \cite{Bur03}. 
Using these figures we find for the radius of the neutron star 
R\,=\,(R$_{\mbox{opt}}^2$\,+\,R$_{\mbox{x}}^2$)$^{1/2}$\,$>$\,16.5\,km (3\,$\sigma$). 

As an alternative we discuss a model with a continuous temperature 
distribution (c.f. Fig.\,\ref{fig:1}b of the form 
\begin{equation}
\mbox{T} = \mbox{T}_{\mbox{hot}} \times  {\left[ 1+\left( \frac{\theta}{\theta_0}\right) ^{\gamma} \right]}^{-1}
\end{equation}
The best fit to both the optical and the X-ray spectrum yields a 
central temperature of the hot spot T$_{\mbox{hot}}$\,=\,82\,eV, an angular size 
of the hot spot $\theta_0$\,=\,40$^{\circ}$ and $\gamma$\,=\,2.1. 
In this case the neutron star radius is 16.8\,km. 
We note that such hot spots may be caused by an anisotropic heat 
transport in strong magnetic fields ($>$\,10$^{13}$\,G \cite{Gre83}) 
or by polar cap heating in a millisecond pulsar \cite{Pav03}.

The quoted apparent radii R\,=\,16.5\,km for the two-component model 
(Fig.\,\ref{fig:1}a) and R\,=\,16.8\,km for the continuous temperature 
distribution model (Fig.\,\ref{fig:1}b) represent rather lower limits since 
they have been derived under the assumption of blackbody emission. 
These apparent radii R are related to the true stellar radius 
R$_0$ which is usually quoted in the literature by 
\begin{equation}
\mbox{R} = {\mbox{R}}_0 (1-{\mbox{R}}_{\mbox{s}}/{\mbox{R}}_0)^{-1/2} 
\end{equation}
where R$_{\mbox{s}}$\,=\,2GM/c$^2$ is the Schwarzschild radius. 
For a standard neutron star of 1.4 solar masses the true 
radii are R$_0$\,=\,14.0\,km (Fig.\,\ref{fig:1}a) and
R$_0$\,=\,14.1\,km (Fig.\,\ref{fig:1}b), 
respectively, and thus considerably larger than the usual standard 
radius of 10\,km. This implies a rather stiff equation of state. 
We note that the same conclusion was reached by Braje \& Romani\,\cite{Bra02} 
using a two component model and similar arguments. 
In order to compare our results in more detail with the predictions 
of theoretical neutron star models we use the mass-radius diagram 
given by Pons et al.\,\cite{Pon02}. 
This diagram is shown in Fig.\,\ref{fig:2} to which we have added 
a curve corresponding to the apparent radius derived in this paper. 
It is evident that the lower limit  represented by the thick 
dashed curve excludes the quark star discussed in \cite{Pon02} 
with high significance. 
It even excludes the quark star models discussed by 
Schertler et al.\,\cite{Sch98} which have even smaller radii, 
as well as neutron stars with strange quark matter cores 
discussed in the same paper. 
In view of the small uncertainties in the distance 
(117\,$\pm$\,12)\,pc of RXJ1856 and the fact that the blackbody 
leads to an underestimate of the emitting area we conclude 
that this object is a neutron star with a stiff equation 
of state and that the possibility for a quark star and a 
quark core star can be ruled out with high confidence.

\end{document}